\documentclass[12pt]{article}

\usepackage{graphicx}
\usepackage{amsmath,amssymb}
\usepackage{bm}
\usepackage{graphicx, color}
\usepackage{wrapfig}
\begin{document}
\title{
\begin{flushright}
\ \\*[-80pt] 
\begin{minipage}{0.2\linewidth}
\normalsize
\end{minipage}
\end{flushright}
{\Large \bf 
$S_4$ Flavor Symmetry of Quarks and Leptons \\
in SU(5) GUT
\\*[20pt]}}

\author{
\centerline{
Hajime~Ishimori$^{1,}$\footnote{E-mail address: ishimori@muse.sc.niigata-u.ac.jp},   
~Yusuke~Shimizu$^{1,}$\footnote{E-mail address: shimizu@muse.sc.niigata-u.ac.jp}, } \\ 
\centerline{
Morimitsu~Tanimoto$^{2,}$\footnote{E-mail address: tanimoto@muse.sc.niigata-u.ac.jp} }
\\*[20pt]
\centerline{
\begin{minipage}{\linewidth}
\begin{center}
$^1${\it \normalsize
Graduate~School~of~Science~and~Technology,~Niigata~University, \\ 
Niigata~950-2181,~Japan } \\
$^2${\it \normalsize
Department of Physics, Niigata University,~Niigata 950-2181, Japan } 
\end{center}
\end{minipage}}
\\*[50pt]}

\date{
\centerline{\small \bf Abstract}
\begin{minipage}{0.9\linewidth}
\medskip 
\medskip 
\small
 We  present  a  $S_4$ flavor model to unify
  quarks and leptons in the framework of the SU(5) GUT.
 Three generations of $\overline 5$-plets in SU(5) are assigned $3_1$
of $S_4$ while the  first and the second generations of 
$10$-plets  in  SU(5)  are assigned to be  $2$ of $S_4$,
and the third generation of $10$-plet is to be  $1_1$ of $S_4$.
Right-handed neutrinos are also assigned  
 $2$ for the first and second generations
and $1_1$ for  the third generation, respectively.
 Taking vacuum alignments of  relevant gauge singlet scalars,
 we predict  the  quark  mixing  as well as the tri-bimaximal
mixing of neutrino flavors. Especially, the Cabbibo angle is
predicted to be $15^{\circ}$ in the limit of  the vacuum alignment.
We can improve the model to predict  observed CKM mixing angles
as well as the non-vanishing $U_{e3}$ of the neutrino flavor mixing.
\end{minipage}
}

\begin{titlepage}
\maketitle
\thispagestyle{empty}
\end{titlepage}

\section{Introduction}

Neutrino experimental  data provide  an important clue for elucidating the 
origin of the observed hierarchies in mass matrices for quarks and leptons.
Recent experiments of the neutrino oscillation 
go into a  new  phase  of precise  determination of
 mixing angles and mass squared  differences  \cite{Threeflavors}, 
which indicate the tri-bimaximal mixing  for three flavors 
 in the lepton sector  \cite{HPS}. 
These large  mixing angles are completely  
 different from the quark mixing ones.
Therefore, it is very important
to find a natural model that leads to these mixing patterns
 of quarks and leptons with good accuracy.

 Flavor symmetry is expected to explain  the mass spectrum  and
the mixing matrix of quarks and leptons. 
In particular,  some predictive models with  non-Abelian  discrete flavor
symmetries have been explored by many authors.
Among them,
 the tri-bimaximal mixing of leptons has been   understood 
based on the non-Abelian finite group $S_3$ \cite{S3-FTY}-\cite{S3-Lavoura}, 
$A_4$ \cite{A4-Ma and Rajasekaran}-\cite{A4-Hirsch3}, and 
$T'$ \cite{T-Feruglio}-\cite{T-Ding},
because   these  symmetries provide the definite meaning of generations
 and connects different generations.
On the other hand, much  attention has been devoted to the question 
whether these  models  can be  extended to
describe the observed pattern of quark masses and mixing angles,
and whether these can be made compatible with the SU(5) or SO(10) 
grand unified theory (GUT).

Recently,   group-theoretical arguments indicate that
the discrete symmetry   $S_4$ is the minimal flavor
symmetry compatible with the tri-bimaximal neutrino mixing \cite{Lam}.
Actually, the exact  tri-bimaximal neutrino mixing is realized
 in the $S_4$ flavor model \cite{S4-Morisi}.
Thus, the $S_4$ flavor model is attractive for the lepton sector
\cite{S4-Ma}-\cite{S4-Koide}.
Although an  attempt to unify the quark and lepton sector was presented
towards a grand unified theory of flavor \cite{S4-Hagedorn}, 
  mixing angles are not predicted clearly.

In our work, we present a  $S_4$ flavor model to unify
 the quarks and leptons in the framework of the SU(5) GUT.
The group $S_4$ has irreducible representations 
$1_1,~1_2,~2,~3_1,$ and $3_2$.
 Three generations of $\overline 5$-plets in SU(5) are assigned $3_1$
of $S_4$ while the  first and the second generations of 
$10$-plets  in  SU(5)  are assigned to be  $2$ of $S_4$,
and the third generation of $10$-plet is to be  $1_1$ of $S_4$.
These  assignments of $S_4$ for $\overline 5$ and $10$ 
lead to the  completely different structure 
of  quark and lepton mass matrices.
Right-handed neutrinos, which are SU(5) gauge singlets, 
are also assigned $2$ for the first and second generations,
and $1_1$ for  the third generation, respectively.
These  assignments are  essential to realize the tri-bimaximal mixing
of neutrino flavors.
 Taking  vacuum alignments of relevant gauge singlet scalars,
 we  predict the quark  mixing  as well as the tri-bimaximal
mixing of leptons. Especially, the Cabbibo angle is
predicted to be $15^{\circ }$ in the limit of  the vacuum alignment.
We improve the model to predict  observed CKM mixing angles
 as well as the non-vanishing $U_{e3}$ of the neutrino flavor mixing.

The paper is organized as follows.
We present the prototype of the $S_4$ flavor model of
 quarks and leptons in  SU(5) GUT in section 2, and 
 discuss the lepton sector in section 3, and the quark sector in section 4.
In section 5, we present the improved model with additional scalar
in order to study detail of the model.
Section 6 is devoted to the summary.
In the appendix,  we present the multiplication rules of $S_4$, and  
the scalar  potential analysis.
Vacuum alignments and magnitude of VEVs  are also summarized 
in the appendix.

\section{Prototype of  $S_4$ flavor model in SU(5) GUT}

We present the prototype of  the  $S_4$ flavor model 
in the SU(5) GUT to understand the essence of our model clearly.
We consider  the supersymmetric GUT based on  SU(5).
The flavor symmetry of quarks and leptons is the discrete group $S_4$
 in our model.
The group $S_4$ has irreducible representations $1_1,~1_2,~2,~3_1,$ and $3_2$. The multiplication rules of  $S_4$ are summarized in appendix.

\begin{table}[h]
\begin{tabular}{|c|ccccc||cc|}
\hline
&$T_3$ & $(T_1,T_2)$ & $( F_1, F_2, F_3)$ & $(N_e^c,N_\mu ^c)$ & $N_\tau ^c$ & $H_5$ &$H_{\bar 5} $ \\ \hline
SU(5) & $10$ & $10$ & $\bar 5$ & $1$ & $1$ & $5$ & $\bar 5$ \\
$S_4$ & $1_1$ & $2$ & $3_1$ & $2$ & $1_1$ & $1_1$ & $1_1$ \\
$Z_4$ & $\omega ^2$ & $\omega ^3$ & $\omega $ & $1$ & $1$ & $1$ & $1$ \\
\hline
\end{tabular}
\end{table}
\vspace{-0.5cm}
\begin{table}[h]
\begin{tabular}{|c|cccccc|}
\hline
& $\chi _1$&  $(\chi _2,\chi _3)$ &  $(\chi _4,\chi _5)$&$(\chi _6,\chi _7,\chi _8)$  & $(\chi _9,\chi _{10},\chi _{11})$ & $(\chi _{12},\chi _{13},\chi _{14})$ \\ \hline
SU(5) & $1$ & $1$ & $1$ & $1$ & $1$ & $1$ \\
$S_4$ & $1_1$ & $2$ & $2$ & $3_1$ & $3_1$ & $3_1$ \\
$Z_4$ & $\omega^2$ & $\omega^2$ & $1$ & $\omega ^3$ & $1$ & $\omega $ \\
\hline
\end{tabular}
\caption{Assignments of SU(5), $S_4$, and $Z_4$ representations, 
where  the phase factor $\omega$ is $i$.}
\end{table}

Let us present the model of the quark and lepton  flavor 
with the $S_4$ group in SU(5) GUT. 
In SU(5), matter fields are unified into a $10(q_1,u^c,e^c)_L$ 
and a $\bar 5(d^c,l_e)_L$ dimensional representations. 
Three generations of $\bar 5$, which are denoted by $F_i$,
 are assigned by $3_1$ of $S_4$. 
On the other hand, the third generation of the $10$-dimensional 
representation is assigned by $1_1$ of $S_4$, 
so that the top quark Yukawa coupling is allowed in tree level. 
While, the first and the second generations are assigned $2$ of $S_4$. 
These $10$-dimensional representations are denoted by 
$T_3$ and $(T_1,T_2)$, respectively.
Right-handed neutrinos, which are SU(5) gauge singlets,
are also assigned  $1_1$ and  $2$ for $N^c_\tau$ and  
$(N^c_e,N^c_\mu)$, respectively
\footnote{The similar assignments of right-handed neutrinos were
  presented in the first version of Ref.\cite{S4-Morisi}.}.
 
We introduce new scalars 
$\chi_i$ in addition to the $5$-dimensional 
and $\bar 5$-dimensional Higgs of the SU(5), $H_5$ and $H_{\bar 5} $
 which are  assigned $1_1$ of $S_4$. 
These new scalars are supposed to be SU(5) gauge singlets. 
The  $\chi_1$ scalar is assigned $1_1$, $(\chi _2,\chi _3)$, 
and $(\chi _4,\chi _5)$ are $2$, $(\chi _6,\chi _7,\chi _8),~(\chi _9,\chi _{10},\chi _{11})$, and $(\chi _{12},\chi _{13},\chi _{14})$ 
are $3_1$ of the $S_4$ representations, respectively. 
The  $\chi_1$ and $(\chi _2,\chi _3)$  scalars
are coupled with the  up type  quark sector, $(\chi _4,\chi _5)$ are  
coupled with the right-handed Majorana neutrino sector, 
$(\chi _6,\chi _7,\chi _8)$ are  coupled with the Dirac neutrino sector, 
$(\chi _9,\chi _{10},\chi _{11})$ and $(\chi _{12},\chi _{13},\chi _{14})$ 
are coupled with the charged lepton and down type quark sector, respectively. 
We also add $Z_4$ symmetry in order to obtain   relevant couplings.
The particle assignments of SU(5), $S_4$, and $Z_4$ are summarized Table 1.

 We can now write down  the superpotential at the leading order 
 in terms of the cut off scale $\Lambda$, 
which is taken to be the Planck scale. The SU(5) invariant superpotential 
of the Yukawa  sector respecting  $S_4$ and $Z_4$ symmetries is given as
\begin{align}
w_\text{SU(5)}^{(0)} &= y_1^u(T_1,T_2)\otimes (T_1,T_2)\otimes \chi _1\otimes H_5/\Lambda + y_2^u(T_1,T_2)\otimes (T_1,T_2)\otimes (\chi _2,\chi _3)\otimes H_5/\Lambda \nonumber \\
&\ + y_3^uT_3\otimes T_3\otimes H_5 + M_1(N_e^c,N_\mu ^c)\otimes (N_e^c,N_\mu ^c) + M_2N_\tau ^c\otimes N_\tau ^c \nonumber \\
&\ + y^N(N_e^c,N_\mu ^c)\otimes (N_e^c,N_\mu ^c)\otimes (\chi _4,\chi _5) \nonumber \\
&\ + y_1^D(N_e^c,N_\mu ^c)\otimes (F_1,F_2,F_3)\otimes (\chi _6,\chi _7,\chi _8)\otimes H_5/\Lambda \nonumber \\
&\ + y_2^DN_\tau ^c\otimes (F_1,F_2,F_3)\otimes (\chi _6,\chi _7,\chi _8)\otimes H_5/\Lambda \nonumber \\
&\ + y_1(F_1,F_2,F_3)\otimes (T_1,T_2)\otimes (\chi _9,\chi _{10},\chi _{11})\otimes H_{\bar 5}/\Lambda  \nonumber \\
&\ + y_2(F_1,F_2,F_3)\otimes T_3\otimes (\chi _{12},\chi _{13},\chi _{14})\otimes H_{\bar 5}/\Lambda ,
\end{align}
 where  $M_1$ and $M_2$ are mass parameters
 for right-handed Majorana neutrinos,
and  Yukawa coupling constants $y_i^a$ and $y_i$ are complex in general.
By decomposing this superpotential into the quark sector and 
the lepton sector,
we can discuss mass matrices of quarks and leptons in  following sections.


\section{Lepton sector}
At first, we begin to discuss the  lepton sector of 
the superpotential $w_\text{SU(5)}^{(0)}$.
 Denoting Higgs doublets as $h_u$ and $h_d$,
 the superpotential of the Yukawa sector respecting 
the $S_4 \times Z_4$ symmetry  is given for charged leptons as
\begin{align}
w_l &= y_1\left [\frac{e^c}{\sqrt2}(l_\mu \chi _{10}-l_\tau \chi _{11})+\frac{\mu ^c}{\sqrt 6}(-2l_e \chi _{9}+l_\mu \chi _{10}+l_\tau \chi _{11})
\right ] h_d/\Lambda \nonumber \\
&\ + y_2\tau ^c ( l_e \chi _{12}+l_\mu \chi _{13}+l_\tau \chi _{14})h_d/\Lambda .
\end{align}
For  right-handed Majorana neutrinos,  the superpotential is given as
\begin{align}
w_N &= M_1(N_e^cN_e^c+N_\mu ^cN_\mu ^c) + M_2N_\tau ^cN_\tau ^c \nonumber \\
&\ +y^N\left[(N_e^cN_\mu ^c+N_\mu ^cN_e^c)\chi _4+(N_e^cN_e^c-N_\mu ^cN_\mu ^c)\chi _5\right ],
\end{align}
and for  neutrino Yukawa couplings,  the superpotential is
\begin{align}
w_D &= y_1^D\left [\frac{N_e^c}{\sqrt2}(l_\mu \chi _7-l_\tau \chi _8) +\frac{N_\mu ^c}{\sqrt 6}(-2l_e \chi _6 +l_\mu \chi _7+l_\tau \chi _8)\right ]
 h_u/\Lambda \nonumber \\
&\ +y_2^DN_\tau ^c(l_e\chi _6+l_\mu \chi _7+l_\tau \chi _{8})h_u/\Lambda .
\end{align}

 Higgs doublets $h_u,h_d$ and gauge singlet scalars $\chi _i$, 
are assumed to develop their vacuum expectation values (VEVs) as follows:
\begin{align}
&\langle h_u\rangle =v_u,
\quad
\langle h_d\rangle =v_d,
\quad
\langle (\chi _4,\chi _5)\rangle =(u_4,u_5),
\quad
\langle (\chi _6,\chi _7,\chi _8)\rangle =(u_6,u_7,u_8), \nonumber \\
&\langle (\chi _9,\chi _{10},\chi _{11})\rangle =(u_9,u_{10},u_{11}),
\quad 
\langle (\chi _{12},\chi _{13},\chi _{14})\rangle =(u_{12},u_{13},u_{14}),
\end{align}
which are supposed to be real.
Then, we obtain the mass matrix for charged leptons as
\begin{equation}
M_l = y_1v_d\begin{pmatrix}
                          0 & \alpha _{10}/\sqrt 2 & -\alpha _{11}/\sqrt 2 \\
                          -2\alpha _9/\sqrt 6 & \alpha _{10}/\sqrt 6 & \alpha _{11}/\sqrt 6   \\
                          0 & 0 & 0 \end{pmatrix}
+y_2v_d\begin{pmatrix}
               0 & 0 & 0 \\
               0 & 0 & 0 \\
               \alpha _{12} & \alpha _{13} & \alpha _{14} \\
          \end{pmatrix},
\label{charged}
\end{equation}
while the right-handed Majorana neutrino mass matrix is given as
\begin{equation}
M_N = \begin{pmatrix}
               M_1+y^N\alpha _5\Lambda & y^N\alpha _4\Lambda & 0 \\
               y^N\alpha _4\Lambda & M_1-y^N\alpha _5\Lambda & 0 \\
               0 & 0 & M_2
         \end{pmatrix},
\label{majorana}
\end{equation}
and the Dirac mass matrix of neutrinos is
\begin{equation}
M_D = y_1^Dv_u\begin{pmatrix}
                          0 & \alpha _7/\sqrt 2 & -\alpha _8/\sqrt 2 \\
                          -2\alpha _6/\sqrt 6 & \alpha _7/\sqrt 6 & \alpha _8/\sqrt 6   \\
                          0 & 0 & 0 \end{pmatrix}+y_2^Dv_u\begin{pmatrix}
                                                          0 & 0 & 0 \\
                                                          0 & 0 & 0 \\
                                        \alpha _6 & \alpha _7 & \alpha _8
                                                     \end{pmatrix},
\label{dirac}
\end{equation}
where we denote $\alpha_i \equiv  u_i/\Lambda$.


Let us discuss  lepton masses and mixing angles by considering mass matrices
in   Eqs.(\ref{charged}), (\ref{majorana}) and (\ref{dirac}).
In order to get the left-handed mixing of  charged leptons,
we investigate $M_l^\dagger M_l$:
\begin{align}
&M_l^\dagger M_l = v_d^2\times \nonumber \\
&\begin{pmatrix}
\frac{2}{3}|y_1|^2\alpha _9^2+|y_2|^2\alpha _{12}^2 & -\frac{1}{3}|y_1|^2\alpha _9\alpha _{10}+ |y_2|^2{\alpha _{12}}\alpha _{13} & -\frac{1}{3}|y_1|^2\alpha _9\alpha _{11}+|y_2|^2\alpha _{12}\alpha _{14} \\
-\frac{1}{3}|y_1|^2\alpha _9\alpha _{10}+|y_2|^2{\alpha _{12}}\alpha _{13} & \frac{2}{3}|y_1|^2\alpha _{10}^2+|y_2|^2\alpha _{13}^2 & -\frac{1}{3}|y_1|^2\alpha _{10}\alpha _{11}+|y_2|^2\alpha _{13}\alpha _{14} \\
-\frac{1}{3}|y_1|^2\alpha _9\alpha _{11}+|y_2|^2\alpha _{12}\alpha _{14} & -\frac{1}{3}|y_1|^2\alpha _{10}\alpha _{11}+|y_2|^2\alpha _{13}\alpha _{14} & \frac{2}{3}|y_1|^2\alpha _{11}^2+|y_2|^2\alpha _{14}^2
 \end{pmatrix}.
\end{align}
If we can take   vacuum alignments 
$(u_{9}, u_{10},  u_{11})=(u_{9}, u_{10}, 0)$ and 
$(u_{12}, u_{13},  u_{14})=(0,0, u_{14})$, that is
 $\alpha_{11} = \alpha_{12} = \alpha_{13} = 0$,
we obtain 
\begin{equation}
M_l ^\dagger M_l = v_d^2
\begin{pmatrix}
\frac{2}{3}|y_1|^2\alpha _9^2 & -\frac{1}{3}|y_1|^2\alpha _9\alpha _{10} & 0 \\
-\frac{1}{3}|y_1|^2\alpha _9\alpha _{10} & \frac{2}{3}|y_1|^2\alpha _{10}^2 & 0 \\
0 & 0 & |y_2|^2\alpha _{14}^2
 \end{pmatrix},
\end{equation}
which gives
 $\theta^l_{13}=\theta^l_{23}=0$, where $\theta^l_{ij}$ denote
 left-handed mixing angles to diagonalize the charged lepton mass matrix.
Since the electron mass is tiny compared with the muon mass,
 we expect  $\alpha _9 \ll \alpha_{10}$ and then we get the mixing angle
$\theta_{12}^l$ as,
\begin{align}
\tan \theta^l_{12} &\approx -\frac{\alpha _9}{2\alpha _{10}},
\end{align}
 and  charged lepton masses, 
\begin{align}
m_e^2\approx \frac{1}{2}|y_1|^2\alpha _9^2v_d^2 \ ,
\ 
m_\mu ^2\approx 
\frac{2}{3}|y_1|^2\alpha _{10}^2v_d^2+\frac{1}{6}|y_1|^2\alpha _9^2v_d^2\approx \frac{2}{3}|y_1|^2\alpha _{10}^2v_d^2\ ,
\ 
m_\tau ^2=|y_2|^2\alpha _{14}^2v_d^2\ .
\label{chargemass}
\end{align}
Therefore, the mixing of $\theta^l_{12}$ is estimated as
\begin{align}
|\tan \theta^l_{12}|\approx \frac{1}{\sqrt 3}&\frac{m_e}{m_\mu}\approx 
2.8\times 10^{-3},
\label{leptonmix}
\end{align}
which is negligibly small.
The mixing angle $\theta_{13}^l$ is 
at most ${\cal O}(m_e/m_\tau)$
even if we take into account of non-zero $\alpha_{12}$.
These tiny   $\theta_{12}^l$ and $\theta_{13}^l$
hardly affect the magnitude of  the lepton mixing matrix element  $U_{e3}$,
which will be discussed later.

It is noticed that one can  take at the leading order
the vacuum alignment $(u_{9}, u_{10}, u_{11})=(0, u_{10}, 0)$
in order to guarantee  $\alpha_9 \ll \alpha_{10}$,
in which the electron mass vanishes.
 In conclusion, we find the charged lepton mass matrix 
to be almost diagonal one.


 Taking vacuum alignments $(u_{4}, u_{5})=(0, u_{5})$
and   $(u_{6}, u_{7}, u_{8})=(u_{6}, u_{6}, u_{6})$ in Eq.(\ref{majorana}),
 the Majorana mass matrix of neutrinos turns to
\begin{equation}
M_N = \begin{pmatrix}
               M_1+y^N\alpha _5\Lambda & 0 & 0 \\
               0 & M_1-y^N\alpha _5\Lambda & 0 \\
               0 & 0 & M_2
         \end{pmatrix},
\end{equation}
and   the Dirac mass matrix of neutrinos turns to
\begin{equation}
M_D = y_1^Dv_u\begin{pmatrix}
                          0 & \alpha _6/\sqrt 2 & -\alpha _6/\sqrt 2 \\
                          -2\alpha _6/\sqrt 6 & \alpha _6/\sqrt 6 & \alpha _6/\sqrt 6   \\
                          0 & 0 & 0 \end{pmatrix}+y_2^Dv_u\begin{pmatrix}
                                                          0 & 0 & 0 \\
                                                          0 & 0 & 0 \\
                                                          \alpha _6 & \alpha _6 & \alpha _6
                                                     \end{pmatrix}.
\end{equation}
By using the seesaw mechanism $M_\nu = M_D^TM_N^{-1}M_D$, 
the left-handed Majorana neutrino mass matrix is  written as
\begin{equation}
M_\nu = \begin{pmatrix}
                 a+\frac{2}{3}b & a-\frac{1}{3}b & a-\frac{1}{3}b \\
                 a-\frac{1}{3}b & a+\frac{1}{6}b+\frac{1}{2}c & a+\frac{1}{6}b-\frac{1}{2}c \\
                 a-\frac{1}{3}b & a+\frac{1}{6}b-\frac{1}{2}c & a+\frac{1}{6}b+\frac{1}{2}c
            \end{pmatrix},
\label{neutrino}
\end{equation}
where
\begin{equation}
a = \frac{(y_2^D\alpha _6v_u)^2}{M_2},\qquad 
b = \frac{(y_1^D\alpha _6v_u)^2}{M_1-y^N\alpha _5\Lambda },\qquad 
c = \frac{(y_1^D\alpha _6v_u)^2}{M_1+y^N\alpha _5\Lambda }.
\end{equation}
The neutrino mass matrix is decomposed as
\begin{equation}
M_\nu = \frac{b+c}{2}\begin{pmatrix}
                                  1 & 0 & 0 \\
                                  0 & 1 & 0 \\
                                  0 & 0 & 1
                             \end{pmatrix} + \frac{3a-b}{3}\begin{pmatrix}
                                                   1 & 1 & 1 \\
                                                   1 & 1 & 1 \\
                                                   1 & 1 & 1
                \end{pmatrix} + \frac{b-c}{2}\begin{pmatrix}
                                 1 & 0 & 0 \\
                                 0 & 0 & 1 \\
                                 0 & 1 & 0
                                            \end{pmatrix}.
\end{equation}
As well known, the neutrino mass matrix with the tri-bimaximal mixing 
 is expressed in terms of neutrino mass eigenvalues $m_1$, $m_2$ and $m_3 $ as
\begin{equation}
M_\nu = \frac{m_1+m_3}{2}\begin{pmatrix}
                                         1 & 0 & 0 \\ 
                                         0 & 1 & 0 \\
                                         0 & 0 & 1
           \end{pmatrix}+\frac{m_2-m_1}{3}\begin{pmatrix}
                           1 & 1 & 1 \\
                           1 & 1 & 1 \\
                           1 & 1 & 1
       \end{pmatrix}+\frac{m_1-m_3}{2}\begin{pmatrix}
                           1 & 0 & 0 \\
                           0 & 0 & 1 \\
                           0 & 1 & 0
                   \end{pmatrix}.
\end{equation}
Therefore, our neutrino mass matrix $M_\nu $ gives
 the  tri-bimaximal mixing matrix $U_\text{tri-bi}$ and 
  mass eigenvalues  as follows:
\begin{equation}
U_\text{tri-bi} = \begin{pmatrix}
               \frac{2}{\sqrt{6}} &  \frac{1}{\sqrt{3}} & 0 \\
     -\frac{1}{\sqrt{6}} & \frac{1}{\sqrt{3}} &  -\frac{1}{\sqrt{2}} \\
      -\frac{1}{\sqrt{6}} &  \frac{1}{\sqrt{3}} &   \frac{1}{\sqrt{2}}
         \end{pmatrix},
\qquad m_1 = b\ ,\qquad m_2 = 3a\ ,\qquad m_3 = c\ .
\end{equation}
We remind ourselves  that the flavor mixing from the charged lepton sector
 is negligibly small.

  Defining parameters $\mu_0=v_u/\Lambda$,
$\lambda_1=M_1/\Lambda$ and $\lambda_2=M_2/\Lambda$, 
and taking  $y_1^D=y_2^D$,
 the observed values  $\Delta m_{\rm atm}^2$ and 
$\Delta m_{\rm sol}^2$ are expressed as 
\begin{eqnarray}
\Delta m_{\rm atm}^2
=-\frac{4y^N\alpha_5\lambda_1}
{(\lambda_1^2-y^{N2}\alpha_5^2)^2}\ (y_1^D\alpha_6)^4\mu_0^2 v_u^2 ,
\quad
\Delta m_{\rm sol}^2
=\frac{9(\lambda_1^2-y^N\alpha_5)^2-\lambda_2^2}
{\lambda_2^2(\lambda_1-y^N\alpha_5)^2} \ (y_1^D\alpha_6)^4 \mu_0^2v_u^2.
\end{eqnarray}

\begin{wrapfigure}{r}{7.5cm}
\begin{center}
\includegraphics[width=7.5cm]{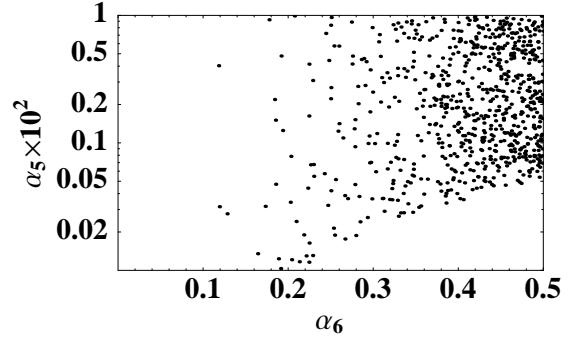}
\caption{The allowed region on the  $\alpha_5-\alpha_6$ plane.}
\end{center}
\end{wrapfigure}

Putting 
 $\Lambda= 2.43\times 10^{18}{\rm GeV}$ and experimental values
of $\Delta m_{\rm atm}^2=(2.1-2.8)\times 10^{-3} {\rm eV^2}$ 
and $\Delta m_{\rm sol}^2=(7.1-8.3)\times 10^{-5} {\rm eV^2}$ 
\cite{Threeflavors}, 
we can estimate  magnitudes of $\alpha_5$ and $\alpha_6$
in the case of the normal neutrino mass hierarchy.
We show the numerical result  in Figure 1, 
where we fix  $y^N=-1$ and $y_1^D=1$.
 We find  $\alpha_6\geq 0.1$, which is much larger than 
 $\alpha_5\approx 10^{-4}-10^{-2}$. 


 Let us discuss  a  possible case of the non-vanishing  $U_{e3}$, which
is the deviation from the tri-bimaximal mixing. 
If $\alpha _4 \not = 0$, which corresponds to the non-vanishing $u_4$,
 the left-handed Majorana neutrino mass matrix deviates
 from Eq.(\ref{neutrino}). 
After rotating by the  tri-bimaximal mixing matrix 
$(\hat M_\nu = U_\text{tri-bi}^\text{T}M_\nu U_\text{tri-bi})$, we obtain
 off-diagonal elements in the neutrino mass matrix
 due to the non-zero  $\alpha_{4}$ as follows:
\begin{equation}
\hat M_\nu = \alpha _6^2v_u^2\begin{pmatrix}
                                               \frac{{y_1^D}^2(M_1+y^N\alpha _5\Lambda )}{M_1^2-{y^N}^2(\alpha _4^2+\alpha _5^2)\Lambda ^2} & 0 & 
-\frac{{y_1^D}^2y^N\alpha _4\Lambda }{M_1^2-{y^N}^2(\alpha _4^2+\alpha _5^2)\Lambda ^2} \\
                                               0 & \frac{3{y_2^D}^2}{M_2} & 0 \\
                           -\frac{{y_1^D}^2y^N\alpha _4\Lambda }{M_1^2-{y^N}^2(\alpha _4^2+\alpha _5^2)\Lambda ^2} & 0 & \frac{{y_1^D}^2(M_1-y^N\alpha _5\Lambda )}{M_1^2-{y^N}^2(\alpha _4^2+\alpha _5^2)\Lambda ^2}
                                           \end{pmatrix}.
\end{equation}
Then the mixing angle  $\delta\theta_{13}^\nu$, 
 which  diagonalizes this mass matrix, is given as 
\begin{equation}
\tan 2\delta \theta _{13}^\nu = \frac{\alpha _4}{\alpha _5},
\end{equation}
which leads to 
\begin{equation}
|U_{e3}|=\frac{2}{\sqrt{6}} |\delta\theta_{13}^\nu | \approx 
\frac{1}{\sqrt{6}} \left |\frac{\alpha_4}{\alpha_5}\right |\ ,
\qquad
|U_{e 2}|=\frac{1}{\sqrt{3}} \ , \qquad
|U_{\mu 3}|\approx \left |\frac{1}{\sqrt{2}}+
\frac{1}{2\sqrt{6}} \frac{\alpha_4}{\alpha_5} \right | \ .
\end{equation}
Thus, the magnitude of $U_{e3}$ is determined by the
non-vanishing ratio $\alpha_4/\alpha_5$.


\section{Quark sector}
In this section, we discuss quark sector of 
the superpotential $w_\text{SU(5)}^{(0)}$.
For up type quarks, the superpotential of the Yukawa sector with
$S_4 \times Z_4$  is given as
\begin{align}
w_u &= y_1^u( u^cq_1+c^cq_2)\chi _1 h_u/\Lambda \nonumber \\
&\ +y_2^u\left[ ( u^cq_2+c^cq_1) \chi _2+(u^cq_1- c^cq_2)\chi _3\right ]
h_u/\Lambda 
 +y_3^u t^c q_3h_u.
\end{align}
For down type quarks, we can write the superpotential as follows:
\begin{align}
w_d &= y_1\left [\frac{ 1}{\sqrt 2}( s^c\chi _{10} - b^c \chi _{11}) q_1+\frac{1}{\sqrt 6}(-2d^c \chi _9+s^c\chi _{10}+b^c\chi _{11})q_2 \right ]
h_d/\Lambda \nonumber \\
&\ + y_2( d^c\chi _{12}+ s^c \chi _{13}+  b^c\chi _{14})q_3 h_d/\Lambda .
\end{align}
We assume 
that scalar fields,  $\chi_i$, develop their VEVs as follows:
\begin{align}
&\langle \chi _1\rangle =u_1,
\quad 
\langle (\chi _2,\chi _3)\rangle =(u_2,u_3),
\nonumber \\
&\langle (\chi _9,\chi _{10},\chi _{11})\rangle =(u_9,u_{10},u_{11}),
\quad
\langle (\chi _{12},\chi _{13},\chi _{14})\rangle =(u_{12},u_{13},u_{14}).
\end{align}
Then,  the mass matrix for up type quarks is given as
\begin{equation}
M_u = v_u\begin{pmatrix}
                  y_1^u\alpha _1+y_2^u\alpha _3 & y_2^u\alpha _2 & 0 \\ 
                  y_2^u\alpha _2 & y_1^u \alpha _1-y_2^u\alpha _3 & 0 \\
                  0 & 0 & y_3^u
             \end{pmatrix},
\end{equation}
and  the down type quark mass matrix is given as
\begin{equation}
M_d = y_1v_d\begin{pmatrix}
                      0 & -2\alpha _9/\sqrt 6 & 0 \\ 
                      \alpha _{10}/\sqrt 2 & \alpha _{10}/\sqrt 6  &  0 \\
                      -\alpha _{11}/\sqrt 2  & \alpha _{11}/\sqrt 6 & 0
                  \end{pmatrix} + y_2v_d\begin{pmatrix}
                                                     0 & 0 & \alpha _{12} \\ 
                                                     0 & 0 & \alpha _{13} \\
                                                     0 & 0 & \alpha _{14}
                                                \end{pmatrix}.
\end{equation}

Let us discuss masses and mixing of the quark sector.
For up type quarks, if we take 
\begin{equation}
\alpha _3 = 0,
\qquad 
y_1^u\alpha _1 = y_2^u\alpha _2,
\label{cond-up}
\end{equation}
which will be  reexamined to get observed  CKM mixing  angles 
 in section 5.2, then, we have
\begin{equation}
M_u = v_u\begin{pmatrix}
                 y_1^u\alpha _1 & y_1^u\alpha _1 & 0 \\
                 y_1^u\alpha _1 & y_1^u\alpha _1 & 0 \\
                 0 & 0 & y_3^u
             \end{pmatrix} ,
\end{equation}
which is diagonalized by the orthogonal matrix $U_u$
\begin{align}
U_u = \begin{pmatrix}
        \cos 45^\circ & \sin 45^\circ & 0 \\
        -\sin 45^\circ & \cos 45^\circ & 0 \\
        0 & 0 & 1
         \end{pmatrix}.
\end{align}
The up type quark masses   are given as 
\begin{align}
m_u=0,
\qquad &m_c=2y_1^uv_u\alpha _1,
\qquad m_t=y_3^uv_u.
\end{align}

For down type quarks, 
putting $\alpha _{11} = \alpha _{12} = \alpha _{13} = 0$,
which is the condition in  the charged lepton sector, we have
\begin{equation}
M_d^\dagger M_d = v_d^2\begin{pmatrix}
                                 \frac{1}{2}|y_1|^2\alpha _{10}^2 & 
\frac{1}{2\sqrt 3}|y_1|^2\alpha _{10}^2 & 0 \\
                                     \frac{1}{2\sqrt 3}|y_1|^2\alpha _{10}^2 & \frac{1}{6}|y_1|^2(4\alpha _9^2+\alpha _{10}^2) & 0 \\
                                     0 & 0 & |y_2|^2\alpha _{14}^2
                                 \end{pmatrix}.
\end{equation}
Then, the mass matrix is  diagonalized  by the orthogonal matrix $U_d$ as
\begin{align}
U_d = \begin{pmatrix}
            \cos 60^\circ & \sin 60^\circ & 0 \\
            -\sin 60^\circ & \cos 60^\circ & 0 \\
            0 & 0 & 1
         \end{pmatrix},
\label{Ud}
\end{align}
where  the small  $\alpha_9$ is neglected. 
The down type quark masses  are given as 
\begin{align}
m_d^2&\approx \frac{1}{2}|y_1|^2\alpha _9^2v_d^2 \ ,
\qquad  
m_s^2\approx \frac{2}{3}|y_1|^2\alpha _{10}^2v_d^2\ ,
\qquad 
m_b^2\approx |y_2|^2\alpha _{14}^2v_d^2\ ,
\end{align}
which are the same ones of  charged lepton masses in Eq.(\ref{chargemass}).

Now, we get  the CKM matrix as follows:
\begin{equation}
V^{CKM} = U_u^\dagger U_d = \begin{pmatrix}
                                      \cos 15^\circ & \sin 15^\circ & 0 \\
                                      -\sin 15^\circ & \cos 15^\circ & 0 \\
                                           0 & 0 & 1
                                       \end{pmatrix}.
\end{equation}
Therefore, in our prototype model of SU(5) GUT with the $S_4$ flavor symmetry,
 the quark sector has a   non-vanishing mixing angle $15^\circ$
 only between the first and second generations
while the lepton flavor mixing is tri-bimaximal.
In order to get the non-vanishing but small mixing angles 
 $V_{cb}^{CKM}$ and $V_{ub}^{CKM}$, we  improve the prototype  model
 in the next section.

\section{Improved $S_4$ flavor model in SU(5) GUT}
We improve the prototype model to get  the observed 
quark and lepton mass spectra  and the CKM mixing matrix.
We introduce
 the  SU(5) $45$-dimensional Higgs $h_{45}$, which is required 
to get the difference between the charged lepton mass spectrum
 and the down type quark mass spectrum. 
Moreover, we add   a $S_4$ doublet
 $(\chi _2^\prime ,\chi _3^\prime )$ and  a $S_4$ triplet
 $(\chi _9^\prime ,\chi _{10}^\prime ,\chi _{11}^\prime )$, 
which are  SU(5) gauge singlet scalars.
These  assignments of SU(5), $S_4$, and $Z_4$ are summarized Table 2.
Since the additional scalars do not contribute  to the neutrino sector,
 the result of the neutrino sector in the prototype model is not changed.
Therefore, we discuss only the  charged lepton sector and the quark sector
 in this section.
\begin{table}[h]
\begin{center}
\begin{tabular}{|c|c||cc|}
\hline
& $h_{45}$ & $(\chi _2^\prime ,\chi _3^\prime )$ & $(\chi _9^\prime ,\chi _{10}^\prime ,\chi _{11}^\prime )$ \\ \hline
SU(5) & $45$ & $1$ & $1$ \\
$S_4$ & $1_1$ & $2$ & $3_1$ \\
$Z_4$ & $\omega^2$ & $\omega^3$ & $\omega ^2$ \\
\hline
\end{tabular}
\caption{Assignments of additional scalars in
SU(5), $S_4$, and $Z_4$ representations.}
\end{center}
\end{table}

The superpotential of the Yukawa sector respecting the SU(5), $S_4$ and $Z_4$ 
symmetries  is given as
\begin{equation}
w_\text{SU(5)} = w_\text{SU(5)}^{(0)}+w_\text{SU(5)}^{(1)},
\end{equation}
where we denote
\begin{align}
w_\text{SU(5)}^{(1)} &= y_4^u(T_1,T_2)\otimes T_3\otimes (\chi _2^\prime ,\chi _3^\prime )\otimes H_5 \nonumber \\
&\ + y_1^\prime (F_1,F_2,F_3)\otimes (T_1,T_2)\otimes (\chi _9^\prime ,\chi _{10}^\prime ,\chi _{11}^\prime )\otimes h_{45}.
\end{align}

\subsection{Improved lepton sector}
Let us discuss the improved  lepton sector of the superpotential 
$w_\text{SU(5)}$.
The superpotential of the charged lepton  sector with
 $S_4\times Z_4$  is given as
\begin{align}
w_l &= y_1\left [\frac{e^c}{\sqrt2}(l_\mu \chi _{10}-l_\tau \chi _{11})+\frac{\mu ^c}{\sqrt 6}(-2l_e \chi _{9}+l_\mu \chi _{10}+l_\tau \chi _{11})\right ]
 h_d/\Lambda \nonumber \\
&\ -3y_1^\prime \left [\frac{e^c}{\sqrt2}(l_\mu \chi _{10}^\prime -l_\tau \chi _{11}^\prime )+\frac{\mu ^c}{\sqrt 6}(-2l_e \chi _9^\prime +l_\mu \chi _{10}^\prime +l_\tau \chi _{11}^\prime )\right ] h_{45}/\Lambda \nonumber \\
&\ + y_2\tau ^c ( l_e \chi _{12}+l_\mu \chi _{13}+l_\tau \chi _{14})h_d/\Lambda .
\end{align}
We denote their VEVs as follows:
\begin{align}
&\langle h_{45}\rangle =v_{45},
\quad
\langle (\chi _9,\chi _{10},\chi _{11})\rangle =(u_9,u_{10},u_{11}), 
\nonumber \\
&\langle (\chi _9^\prime ,\chi _{10}^\prime ,\chi _{11}^\prime )\rangle =
(u_9^\prime ,u_{10}^\prime ,u_{11}^\prime ), \quad
\langle (\chi _{12},\chi _{13},\chi _{14})\rangle =(u_{12},u_{13},u_{14}).
\end{align}
Then, we obtain the mass matrix for charged leptons:
\begin{align}
M_l &= y_1v_d\begin{pmatrix}
                          0 & \alpha _{10}/\sqrt 2 & -\alpha _{11}/\sqrt 2 \\
                          -2\alpha _9/\sqrt 6 & \alpha _{10}/\sqrt 6 & \alpha _{11}/\sqrt 6   \\
                          0 & 0 & 0 
                  \end{pmatrix}+y_2v_d\begin{pmatrix}
                                                   0 & 0 & 0 \\
                                                   0 & 0 & 0 \\
                                                   \alpha _{12} & \alpha _{13} & \alpha _{14} 
                                               \end{pmatrix} \nonumber \\
&\ -3y_1^\prime v_{45}\begin{pmatrix}
                                   0 & \alpha _{10}^\prime /\sqrt 2 & -\alpha _{11}^\prime /\sqrt 2 \\
                                   -2\alpha _9^\prime /\sqrt 6 & \alpha _{10}^\prime /\sqrt 6 & \alpha _{11}^\prime /\sqrt 6   \\
                                   0 & 0 & 0
                                \end{pmatrix},
\end{align}
where we denote $\alpha_i = u_i/\Lambda$ and  
$\alpha _j^\prime = u_j^\prime /\Lambda$.
It is noticed that the third  matrix in the right hand side is 
the additional one
compared with the mass matrix of the prototype model in Eq.(\ref{charged}).

 Masses  and mixing angles of the charged lepton sector are 
  similar to those of  the prototype model in Eqs.(\ref{chargemass}) and
(\ref{leptonmix}). 
If we can take  the vacuum alignments 
$(u_{9}, u_{10},  u_{11})=(u_{9}, u_{10}, 0)$,
$(u_{9}^\prime, u_{10}^\prime,  
u_{11}^\prime)=(u_{9}^\prime, u_{10}^\prime, 0)$  and 
$(u_{12}, u_{13},  u_{14})=(0,0, u_{14})$, that is
 $\alpha_{11} = \alpha_{11}^\prime=\alpha_{12} =\alpha_{13} = 0$,
we obtain charged lepton mass matrix as follow:
\begin{equation}
M_l = v_d\begin{pmatrix}
            0 & (y_1\alpha _{10}-3\bar y_1\alpha _{10}^\prime )/\sqrt 2 & 0 \\
            -2(y_1\alpha _9-3\bar y_1\alpha _9^\prime )/\sqrt 6 & (y_1\alpha _{10}-3\bar y_1\alpha _{10}^\prime )/\sqrt 6 & 0 \\
            0 & 0 & y_2\alpha _{14}
         \end{pmatrix},
\end{equation}
where we replace  $y_1^\prime v_{45}$ with $\bar y_1v_d$. 
Since  we have
\begin{eqnarray}
&&M_l ^\dagger M_l= v_d^2 \times\nonumber \\
&&\begin{pmatrix}
\frac{2}{3}|y_1\alpha _9-3\bar y_1\alpha _9^\prime |^2 & -\frac{1}{3}(y_1^\ast \alpha _9-3\bar y_1^\ast \alpha _9^\prime )(y_1\alpha _{10}-3\bar y_1\alpha _{10}^\prime ) & 0 \\
-\frac{1}{3}(y_1\alpha _9-3\bar y_1\alpha _9^\prime )(y_1^\ast \alpha _{10}-3\bar y_1^\ast \alpha _{10}^\prime ) & \frac{2}{3}|y_1\alpha _{10}-3\bar y_1\alpha _{10}^\prime |^2 & 0 \\
0 & 0 & |y_2|^2\alpha _{14}^2
 \end{pmatrix}, \nonumber\\
\end{eqnarray}
masses and mixing angles of the charged leptons as follows:
\begin{eqnarray}
\hspace{-1cm}
&&m_e^2\approx\frac{1}{2}|y_1\alpha _9-3\bar y_1\alpha _9^\prime |^2v_d^2,
\quad 
m_\mu ^2\approx
\frac{2}{3}|y_1\alpha _{10}-3\bar y_1\alpha _{10}^\prime |^2v_d^2, 
\quad
m_\tau ^2\approx |y_2|^2\alpha _{14}^2v_d^2, \nonumber\\
&&
|\theta^l_{12}|=\left |-\frac{y_1\alpha _9-3\bar y_1\alpha _9^\prime }{2(y_1\alpha _{10}-3\bar y_1\alpha _{10}^\prime )}\right | 
\approx \frac{1}{\sqrt 3}\frac{m_e}{m_\mu }\approx 2.8\times 10^{-3},
\quad 
\theta^l_{23}=0,
\quad
\theta^l_{13}=0.
\end{eqnarray}
Thus, the charged lepton mass matrix is almost diagonal, 
and so the tri-bimaximal mixing of neutrino flavors
 is also  realized  in this improved model.


\subsection{Improved quark sector}
Let us discuss the quark sector of the superpotential $w_\text{SU(5)}$.
For up type quarks, the superpotential  respecting 
 $S_4 \times Z_4$  is given as
\begin{align}
w_u &= y_1^u( u^cq_1+c^cq_2)\chi _1 h_u/\Lambda \nonumber \\
&\ +y_2^u\left [ ( u^cq_2+c^cq_1) \chi _2+(u^cq_1- c^cq_2)\chi _3\right ]
h_u/\Lambda \nonumber \\
&\ +y_3^u t^c q_3h_u \nonumber \\
&\ +y_4^u\left [(u^c\chi _2^\prime +c^c\chi _3^\prime )q_3+t^c(q_1\chi _2^\prime +q_2\chi _3^\prime )\right ] h_u/\Lambda .
\end{align}
For down type quarks,  we can write the superpotential as follows:
\begin{align}
w_d &= y_1\left [
\frac{ 1}{\sqrt 2}( s^c\chi _{10} - b^c \chi _{11}) q_1+\frac{1}{\sqrt 6}(-2d^c \chi _9+s^c\chi _{10}+b^c\chi _{11})q_2 \right ]h_d/\Lambda \nonumber \\
&\ + y_1^\prime \left [\frac{ 1}{\sqrt 2}( s^c\chi _{10}^\prime  - b^c \chi _{11}^\prime ) q_1+\frac{1}{\sqrt 6}(-2d^c \chi _9^\prime +s^c\chi _{10}\prime +b^c\chi _{11}^\prime )q_2\right ]
v_{45}/\Lambda \nonumber \\
&\ + y_2( d^c\chi _{12}+ s^c \chi _{13}+  b^c\chi _{14})q_3 h_d/\Lambda .
\end{align}
We denote  their VEVs as follows:
\begin{eqnarray}
&&
\langle \chi _1\rangle =u_1,
\quad 
\langle (\chi _2,\chi _3)\rangle =(u_2,u_3),
\quad 
\langle (\chi _2^\prime ,\chi _3^\prime )\rangle =(u_2^\prime ,u_3^\prime ),
\nonumber \\
&&\langle (\chi _9,\chi _{10},\chi _{11})\rangle =(u_9,u_{10},u_{11}),\quad 
\langle (\chi _9^\prime ,\chi _{10}^\prime ,\chi _{11}^\prime )\rangle =(u_9^\prime ,u_{10}^\prime ,u_{11}^\prime ),  \nonumber\\
&&
\langle (\chi _{12},\chi _{13},\chi _{14})\rangle =(u_{12},u_{13},u_{14}).
\end{eqnarray}
Then, we obtain the mass matrix for up type quarks is given as
\begin{equation}
M_u = v_u\begin{pmatrix}
                  y_1^u\alpha _1+y_2^u\alpha _3 & y_2^u\alpha _2 & y_4^u\alpha _2^\prime \\ 
                  y_2^u\alpha _2 & y_1^u \alpha _1-y_2^u\alpha _3 & y_4^u\alpha _3^\prime \\
                  y_4^u\alpha _2^\prime & y_4^u\alpha _3^\prime & y_3^u
             \end{pmatrix},
\label{Mu}
\end{equation}
while the  down type quark mass matrix is given as
\begin{align}
M_d &= y_1v_d\begin{pmatrix}
                      0 & -2\alpha _9/\sqrt 6 & 0 \\ 
                      \alpha _{10}/\sqrt 2 & \alpha _{10}/\sqrt 6  &  0 \\
                      -\alpha _{11}/\sqrt 2  & \alpha _{11}/\sqrt 6 & 0
                   \end{pmatrix} + y_2v_d\begin{pmatrix}
                                                     0 & 0 & \alpha _{12} \\ 
                                                     0 & 0 & \alpha _{13} \\
                                                     0 & 0 & \alpha _{14}
                                                 \end{pmatrix} \nonumber \\
&\ + y_1^\prime v_{45}\begin{pmatrix}
                                   0 & -2\alpha _9^\prime /\sqrt 6 & 0 \\ 
                                   \alpha _{10}^\prime /\sqrt 2 & \alpha _{10}^\prime /\sqrt 6  &  0 \\
                                  -\alpha _{11}^\prime /\sqrt 2  & \alpha _{11}^\prime /\sqrt 6 & 0
                               \end{pmatrix}.
\label{Md}
\end{align}

We consider the quark  mixing.
The up type quark mass matrix (\ref{Mu})
turns to the following one 
after rotating by $\theta_{12}^u=45^\circ$:
\begin{equation}
\hat M_u = v_u\begin{pmatrix}
                y_1^u\alpha _1-y_2^u\alpha _2 & y_2^u\alpha _3 & \frac{y_4^u}{\sqrt 2}(\alpha _2^\prime -\alpha _3^\prime ) \\
                y_2^u\alpha _3 & y_1^u\alpha _1+y_2^u\alpha _2 & \frac{y_4^u}{\sqrt 2}(\alpha _2^\prime +\alpha _3^\prime ) \\
                \frac{y_4^u}{\sqrt 2}(\alpha _2^\prime -\alpha _3^\prime ) & \frac{y_4^u}{\sqrt 2}(\alpha _2^\prime +\alpha _3^\prime ) & y_3^u
             \end{pmatrix}.
\end{equation}
In order to obtain the non-vanishing quark mixing of 
$V^{CKM}_{cb}$ and $V^{CKM}_{ub}$, we take
\begin{equation}
y_2^u\alpha _3\gg y_1^u\alpha _1 , \ y_2^u\alpha _2,
\qquad 
\alpha _2^\prime =\alpha _3^\prime ,
\end{equation}
which are realized by  vacuum alignments $u_{1}=0$
\footnote{One may consider to remove  
$\chi_1$, which is $S_4$ singlet, in our scheme.},
 $(u_{2}, u_{3})=(0, u_{3})$
and   $(u_{2}^\prime, u_{3}^\prime)=(u_{2}^\prime, u_{2}^\prime)$.
This situation of VEVs is completely different from 
that of the prototype model as seen in Eq.(\ref{cond-up}),
in which $V^{CKM}_{cb}$ and $V^{CKM}_{ub}$ vanish.
Then, we obtain  the so-called Fritzsch-type mass matrix  \cite{Fritzsch}
\begin{equation}
\hat M_u \simeq v_u\begin{pmatrix}
                 0 & y_2^u\alpha _3 & 0 \\
                 y_2^u\alpha _3 & 0 & \sqrt 2y_4^u\alpha _2^\prime \\
                 0 & \sqrt 2y_4^u\alpha _2^\prime & y_3^u
             \end{pmatrix}.
\label{Fritzsch}
\end{equation}
As well known,
the complex phases in this $3\times 3$  matrix  can be removed 
by the phase matrix $P$  as  $P^\dagger \hat M_u P$;
\begin{equation}
  P =\begin{pmatrix}
     1 & 0& 0 \\ 0& e^{-i\rho}& 0\\ 0 & 0 & e^{-i\sigma}
             \end{pmatrix}.
\end{equation}
Therefore, up type quark masses are
\begin{equation}
m_u=\left |\frac{y_3^u{y_2^u}^2\alpha _3^2}{2{y_4^u}^2{\alpha _2^\prime }^2}
\right | v_u ,
\qquad 
m_c=\left |-\frac{2{y_4^u}^2}{y_3^u}{\alpha _2^\prime }^2 \right |v_u
\qquad 
m_t=|y_3^u| v_u,
\end{equation}
and the mixing  matrix to diagonalize $\hat M_u$ in
 Eq.(\ref{Fritzsch}), 
$V_\text{F}~(M_u^\text{diagonal}=V_\text{F}^\dagger \hat M_uV_\text{F})$, is
\begin{equation}
V_\text{F}\approx \begin{pmatrix}
                   1 & \sqrt \frac{m_u}{m_c} & -\sqrt \frac{m_u}{m_t} \\
                  -\sqrt \frac{m_u}{m_c} & 1 & \sqrt \frac{m_c}{m_t} \\
                      \sqrt \frac{m_u}{m_t} & -\sqrt \frac{m_c}{m_t} & 1
                     \end{pmatrix}.
\end{equation}

The conditions from the lepton sector 
$\alpha_{11}=\alpha_{11}^\prime= \alpha_{12}=\alpha_{13}=0$ give
the down type quark mass matrix:
\begin{equation}
M_d = v_d\begin{pmatrix}
            0 & -2(y_1\alpha _9+\bar y_1\alpha _9^\prime )/\sqrt 6 & 0 \\ 
            (y_1\alpha _{10}+\bar y_1\alpha _{10}^\prime )/\sqrt 2 & (y_1\alpha _{10}+\bar y_1\alpha _{10}^\prime )/\sqrt 6 & 0 \\
            0 & 0 & y_2\alpha _{14}
         \end{pmatrix},
\end{equation}
where we denote $\bar y_1v_d=y_1^\prime v_{45}$.
Then, we have
\begin{equation}
M_d^\dagger M_d = v_d^2\begin{pmatrix}
                          \frac{1}{2}|y_1\alpha _{10}+\bar y_1\alpha _{10}^\prime |^2 & \frac{1}{2\sqrt 3}|y_1\alpha _{10}+\bar y_1\alpha _{10}^\prime |^2 & 0 \\
                          \frac{1}{2\sqrt 3}|y_1\alpha _{10}+\bar y_1\alpha _{10}^\prime |^2 & \frac{1}{6}(4|y_1\alpha _9+\bar y_1\alpha _9^\prime |^2+|y_1\alpha _{10}+
\bar y_1\alpha _{10}^\prime |^2) & 0 \\
                          0 & 0 & |y_2|^2\alpha _{14}^2
                       \end{pmatrix}.
\end{equation}
After rotating by $U_d$ in Eq.(\ref{Ud}), $M_d^\dagger M_d$ turns to be
\begin{equation}
 v_d^2\begin{pmatrix}
         \frac{1}{2}|y_1\alpha _9+\bar y_1\alpha _9^\prime |^2 & -\frac{1}{2\sqrt 3}|y_1\alpha _9+\bar y_1\alpha _9^\prime |^2 & 0 \\
         -\frac{1}{2\sqrt 3}|y_1\alpha _9+\bar y_1\alpha _9^\prime |^2 & \frac{1}{6}(|y_1\alpha _9+\bar y_1\alpha _9^\prime |^2+4|y_1\alpha _{10}+\bar y_1\alpha _{10}^\prime |^2) & 0 \\
         0 & 0 & |y_2|^2\alpha _{14}^2
      \end{pmatrix}.
\end{equation}
Then, down type quark masses are  given as
\begin{align}
&m_d^2\approx \frac{1}{2}|y_1\alpha _9+\bar y_1\alpha _9^\prime |^2v_d^2\ ,
\quad  
m_s^2\approx
\frac{2}{3}|y_1\alpha _{10}+\bar y_1\alpha _{10}^\prime |^2v_d^2\ ,
\quad 
m_b^2\approx
 |y_2|^2\alpha _{14}^2v_d^2\ , 
\end{align}
and the mixing angle $\theta_{12}^d$ is  
$60^\circ+\delta\theta_{12}^d$, where
\begin{equation}
\delta \theta _{12}^d=-\frac{\sqrt 3|y_1\alpha _9+\bar y_1\alpha _9^\prime |^2}{4|y_1\alpha _{10}+\bar y_1\alpha _{10}^\prime |^2}=-\frac{m_d^2}{\sqrt 3m_s^2}\approx -1.5\times 10^{-3} .
\end{equation}
Therefore,  $\theta _{12}^d$ is almost $60^\circ$.
We add comments on $\theta_{23}^d$ and  $\theta_{13}^d$,
which vanish in our scheme
because of $\alpha_{11}=\alpha_{11}^\prime= \alpha_{12}=\alpha_{13}=0$.
Non-vanishing   $\alpha_{ij}$ lead to 
\begin{equation}
\theta_{13}^d \approx -\frac{y_1\alpha _{11}}{\sqrt 2y_2\alpha _{14}},
\qquad
\theta _{23}^d \approx \frac{2(y_1\alpha _{10}+\bar y_1\alpha _{10}^\prime )\alpha _{13}-y_1\alpha _{11}\alpha _{14}}{\sqrt 6y_2\alpha _{14}^2},
\label{dmix}
\end{equation}
where the complex phases of Yukawa couplings are neglected.
These mixing angles are expected to be tiny 
as far as $\alpha_{14}\gg \alpha_{11},  \ \alpha_{13}$.

Let us  discuss  the CKM matrix.
The unitary matrices diagonalizing the up type quark mass matrix and 
the down type quark one, 
$U_u$ and $U_d$ are given, respectively,
\begin{eqnarray}
&&U_u = \begin{pmatrix}
                 \cos 45^\circ & \sin 45^\circ & 0 \\
                 -\sin 45^\circ & \cos 45^\circ & 0 \\
                 0 & 0 & 1
             \end{pmatrix}
\begin{pmatrix}
     1 & 0& 0 \\ 0& e^{-i\rho}& 0\\ 0 & 0 & e^{-i\sigma}
             \end{pmatrix}
\begin{pmatrix}  
                  1 & \sqrt \frac{m_u}{m_c} & -\sqrt \frac{m_u}{m_t} \\
                -\sqrt \frac{m_u}{m_c} & 1 & \sqrt \frac{m_c}{m_t} \\
                     \sqrt \frac{m_u}{m_t} & -\sqrt \frac{m_c}{m_t} & 1
                          \end{pmatrix} , 
\nonumber\\
&&U_d = \begin{pmatrix}
                         \cos 60^\circ & \sin 60^\circ & 0 \\
                        -\sin 60^\circ & \cos 60^\circ & 0 \\
                            0 & 0 & 1
                                                 \end{pmatrix}.
\end{eqnarray}
Therefore, the CKM matrix is written as 
\begin{equation}
V^{CKM} = U_u^\dagger U_d =
 \begin{pmatrix}     
      1 & -\sqrt \frac{m_u}{m_c} & \sqrt \frac{m_u}{m_t} \\
      \sqrt \frac{m_u}{m_c} & 1 & -\sqrt \frac{m_c}{m_t} \\
        -\sqrt \frac{m_u}{m_t} & \sqrt \frac{m_c}{m_t} & 1
                                         \end{pmatrix}
\begin{pmatrix}
     1 & 0& 0 \\ 0& e^{i\rho}& 0\\ 0 & 0 & e^{i\sigma}
             \end{pmatrix}
\begin{pmatrix}
                \cos 15^\circ & \sin 15^\circ & 0 \\
                  -\sin 15^\circ & \cos 15^\circ & 0 \\
                     0 & 0 & 1
                \end{pmatrix} . 
\end{equation}
The relevant CKM mixing elements are given as 
\begin{equation}
\left |V_{us}^\text{CKM}\right |=
\left | \sin 15^{\circ} - \cos 15^{\circ}\sqrt{\frac{m_u}{m_c}}e^{i\rho}\right |, \qquad
\left |V_{cb}^\text{CKM}\right |=\sqrt \frac{m_c}{m_t}, \qquad
\left |V_{ub}^\text{CKM}\right |=\sqrt \frac{m_u}{m_t}
.
\end{equation}
In the limit of neglecting the CP violating phase, $\rho=0$,
putting  typical values  at the GUT scale 
 $m_u=1.04\times 10^{-3}$GeV, $m_c=302\times 10^{-3}$ GeV, $m_t=129$GeV,
which are derived in Ref. \cite{Koide},  we predict
\begin{align}
\left |V_{us}^\text{CKM}\right |=0.202, \qquad 
\left |V_{cb}^\text{CKM}\right |=0.048, \qquad
\left |V_{ub}^\text{CKM}\right |=0.003
.
\end{align}
By adjusting the non-zero  phase $\rho=50^\circ$, 
we can get the central value of the observed Cabbibo angle $0.226$.
Another phase $\sigma$ is still a free parameter.
  Our predicted 
 $V_{cb}^\text{CKM}$ and $V_{ub}^\text{CKM}$ are somewhat different
from the central values of observed mixing angles.
If we take into account the down type quark mixing angles in Eq.(\ref{dmix}),
which are neglected in this stage, we expect to improve the situation
 including  the phase $\sigma$.
Then, we can discuss
the CP violating phenomena in the CKM scheme by using 
two phases $\rho$ and $\sigma$.

The difference between  the charged lepton mass spectrum and 
the down type quark  one is occurred by the $45$-dimensional Higgs.
The masses of charged leptons and down type quarks are given as:
\begin{align}
m_e^2&=\frac{1}{2}|y_1\alpha _9-3\bar y_1\alpha _9^\prime |^2v_d^2,\ m_\mu ^2=\frac{2}{3}|y_1\alpha _{10}-3\bar y_1\alpha _{10}^\prime |^2v_d^2,\ \ m_\tau ^2=|y_2|^2\alpha _{14}^2v_d^2, \nonumber \\
m_d^2&=\frac{1}{2}|y_1\alpha _9+\bar y_1\alpha _9^\prime |^2v_d^2,\quad m_s^2=\frac{2}{3}|y_1\alpha _{10}+\bar y_1\alpha _{10}^\prime |^2v_d^2,\quad m_b^2 =|y_2|^2\alpha _{14}^2v_d^2.
\end{align}
In order to get  the  power ratios of masses  
$m_e^2:m_d^2=1:9,\ m_\mu ^2:m_s^2=9:1$,
which is consistent with the observed mass spectra at the GUT scale,
 we require the following conditions:
\begin{align}
&y_1\alpha _9=5\bar y_1\alpha _9^\prime ,
 \quad \text{or} \quad y_1\alpha _9=2\bar y_1\alpha _9^\prime , 
\nonumber \\
&y_1\alpha _{10}=-3\bar y_1\alpha _{10}^\prime ,\quad
 \text{or} \quad y_1\alpha _{10}=0.
\end{align}

\section{Summary}
We have presented  a flavor model with the  $S_4$ symmetry  to unify
  quarks and leptons in the framework of the SU(5) GUT.
 Three generations of $\overline 5$-plets in SU(5) are assigned $3_1$ of 
$S_4$ while the  first and the second generations of 
$10$-plets  in  SU(5)  are assigned to be  $2$ of $S_4$,
and the third generation of $10$-plet is to be $1_1$ of $S_4$.
These  assignments of $S_4$ for $\overline 5$ and $10$ 
lead to the  completely different structure 
of  quark and lepton mass matrices.
Right-handed neutrinos, which are SU(5) gauge singlets, 
are also assigned $2$ for the first and second generations
and $1_1$ for  the third generation, respectively.
These  assignments are  essential to realize the tri-bimaximal mixing
of neutrino flavors.
 Vacuum alignments of scalars are also required to realize the tri-bimaximal 
mixing of neutrino flavors.
Our model predicts the quark  mixing  as well as the tri-bimaximal
mixing of leptons. Especially, the Cabbibo angle is
predicted to be $15^{\circ}$ in the limit of  the vacuum alignment.
We improve the model to predict  observed CKM mixing angles
 as well as the non-vanishing $U_{e3}$ of the neutrino flavor mixing.
The deviation from  $15^{\circ}$ in  $|V_{us}^\text{CKM}|$ is given by
 $\sqrt{m_u/m_c}$, while the non-vanishing  $|V_{cb}^\text{CKM}|$ 
and  $|V_{ub}^\text{CKM}|$ are given by  $\sqrt{m_c/m_t}$
 and   $\sqrt{m_u/m_t}$, respectively.
The non-vanishing $U_{e3}$ of the neutrino flavor mixing
is independent of these deviations.
 We  will discuss the CP violating phenomena in the CKM scheme 
elsewhere by taking account of the corrections of the down type quark sector.

\vspace{0.5cm}
\noindent
{\bf Acknowledgement}

The work of M.T. has been  supported by the
Grant-in-Aid for Science Research
of the Ministry of Education, Science, and Culture of Japan
No. 17540243.

\appendix
\section*{Appendix}

\section{Multiplication rules of $S_4$}
We present the relevant multiplication rules of $S_4$. 
The two-dimensional one  is  given as
\begin{align}
(a_1,a_2 )_{2}\times (b_1,b_2 )_{2}
= (a_1b_1+a_2b_2 )_{1_1}+(-a_1b_2+a_2b_1 )_{1_2}
 +( a_1b_2+a_2b_1,  a_1b_1-a_2b_2 )_{2}\ .
\end{align}
For three-dimensional representation, the product  is  given as
\begin{align}
(a_1,a_2,a_3)_{3_1}\times (b_1,b_2,b_3)_{3_1} &= (a_1b_1+a_2b_2+a_3b_3)_{1_1} \nonumber \\
&\ +\left [
\frac1{\sqrt2}(a_2b_2-a_3b_3),\frac1{\sqrt6}(-2a_1b_1+a_2b_2+a_3b_3)
\right ]_{2} \nonumber \\
&\ +(a_2b_3+a_3b_2,a_1b_3+a_3b_1,a_1b_2+a_2b_1)_{3_1} \nonumber \\
&\ +(a_3b_2-a_2b_3,a_1b_3-a_3b_1,a_2b_1-a_1b_2)_{3_2}\ .
\end{align}
The product of two- and three-dimensional representations is  given as
\begin{align}
(a_1,a_2 )_{2}\times (b_1,b_2,b_3)_{3_1} &= 
\left [a_2b_1,-\frac12(\sqrt3a_1b_2+a_2b_2),\frac12(\sqrt3a_1b_3-a_2b_3)
\right ]_{3_1} \nonumber \\
&\ +\left [a_1b_1,\frac12(\sqrt3a_2b_2-a_1b_2),-\frac12(\sqrt3a_2b_3+a_1b_3)
\right ]_{3_2}\ .
\end{align}

\section{Vacuum alignments and magnitudes of VEVs}
 In our model, we need vacuum alignments of scalar fields $\chi_i$
of $S_4$ doublets and triplets.
 Vacuum alignments   are summarized at the leading order as follows:
\begin{eqnarray}
&& \chi_1=0, \ \  (\chi _2, \chi_3)= (0,1), \ \ 
 (\chi _2^\prime, \chi_3^\prime)= (1,1), \ \ 
 (\chi_4, \chi_5)= (0,1),  \ \ 
 (\chi _6, \chi_7, \chi _8)=(1,1,1) \nonumber \\
&&(\chi _9, \chi_{10}, \chi _{11})=(0,1,0), \quad
(\chi _9^\prime, \chi_{10}^\prime, \chi _{11}^\prime)=(0,1,0), \quad
(\chi _{12}, \chi_{13}, \chi _{14})=(0,0,1), 
\label{alignment}
\end{eqnarray}
where  magnitudes are given in  arbitrary units.
   Non-vanishing  $m_e$ and $m_d$
require tiny deviations from zeros
for $\chi_9$ and $\chi_9^\prime$,
 which could be  realized  in  the next leading order.
 
In order to show  magnitudes of VEVs, we estimate 
  $\alpha_i$ and $\alpha_j^\prime$.
These are given  as follows:
\begin{align}
&\alpha_ 1 = \alpha_ 2 =\alpha_ 4= \alpha _{11} 
= \alpha _{11}^\prime =\alpha _{12}=\alpha _{13}=0, \quad
\alpha_ 2'=\alpha_ 3'=\sqrt{\left |\frac{y_3^um_c}{2{y_4^u}^2v_u}\right |}, 
\quad \alpha_3= \sqrt{\frac{m_u m_c}{{y_2^u}^2v_u^2}},
\nonumber\\
& \alpha_ 5 = \frac{(3{y_2^D}^2m_1-{y_1^D}^2m_2)M_2}{3{y_2^D}^2y^Nm_1\Lambda},
\quad
\alpha_ 6 = \alpha_ 7 = \alpha_ 8 = \frac{\sqrt{m_2M_2}}{\sqrt 3y_2^Dv_u},
\quad \alpha_ 9 = \frac{\sqrt 2(3m_d+m_e)}{4y_1v_d}, \nonumber \\
&\alpha_ 9' = \frac{\sqrt 2(m_d-m_e)}{4\bar{y_1}v_d}, 
\quad \alpha _{10} = \frac{\sqrt 3(3m_s+m_\mu )}{4\sqrt 2y_1v_d},
\quad 
\alpha _{10}' = \frac{\sqrt 3(m_s-m_\mu )}{4\sqrt 2\bar{y_1}v_d}, 
\quad \alpha _{14} = \frac{m_b}{y_2v_d}\ .
\end{align}
Putting typical values of quark masses at the GUT scale \cite{Koide},
$M_2= 10^{16}\text{GeV}$,
and $\tan\beta=3$ ($v_d\approx 55\text{GeV}$,$v_u\approx 165\text{GeV}$)
with taking $1$ for Yukawa couplings,
 we have 
\begin{align}
&\alpha_ 2'\sim 0.03,
\quad 
\alpha _3\sim 1\times 10^{-4},
\quad 
\alpha_5 \sim 10^{-4}-10^{-2},
\quad 
\alpha _6 \geq 0.1,
\quad 
\alpha _9\sim 3\times 10^{-5}, \nonumber \\
&\alpha _9'\sim 7\times 10^{-6}, 
\quad
\alpha_{10}\sim 8\times 10^{-4},
\quad 
\alpha _{10}'\sim 2\times 10^{-4},
\quad 
\alpha _{14}\sim 0.02.
\end{align}

\section{Scalar potential and  vacuum alignments}
We present  the scalar potential to discuss  the vacuum alignment. 
The $SU(5)\times S_4\times Z_4$ invariant superpotential is given as
\begin{align}
w &= \mu _1(\chi _1)_{1_1}^2 + \mu _2(\chi _2,\chi _3)_2^2 +\mu _3(\chi _4,\chi _5)_2^2 \nonumber \\
&\ +\mu _4(\chi _9,\chi _{10},\chi _{11})_{3_1}^2+\mu _5(\chi _9^\prime ,\chi _{10}^\prime ,\chi _{11}^\prime )_{3_1}^2+\mu _6(\chi _6,\chi _7,\chi _8)_{3_1}\otimes (\chi _{12},\chi _{13},\chi _{14})_{3_1} \nonumber \\
&\ + \eta _1(\chi _1)_{1_1}\otimes (\chi _2,\chi _3)_2\otimes (\chi _4,\chi _5)_2 + \eta _2(\chi _1)_{1_1}\otimes (\chi _6,\chi _7,\chi _8)_{3_1}^2 \nonumber \\
&\ + \eta _3(\chi _1)_{1_1}\otimes (\chi _{12},\chi _{13},\chi _{14})_{3_1}^2 + \eta _4(\chi _1)_{1_1}\otimes (\chi _9,\chi _{10},\chi _{11})_{3_1}\otimes (\chi _9^\prime ,\chi _{10}^\prime ,\chi _{11}^\prime )_{3_1} \nonumber \\
&\ + \eta _5(\chi _2,\chi _3)_2^2\otimes (\chi _4,\chi _5)_2 + \eta _6(\chi _2,\chi _3)_2\otimes (\chi _6,\chi _7,\chi _8)_{3_1}^2 \nonumber \\
&\ + \eta _7(\chi _2,\chi _3)_2\otimes (\chi _9,\chi _{10},\chi _{11})_{3_1}\otimes (\chi _9^\prime ,\chi _{10}^\prime ,\chi _{11}^\prime )_{3_1} + \eta _8(\chi _2,\chi _3)_2\otimes (\chi _{12},\chi _{13},\chi _{14})_{3_1}^2 \nonumber \\
&\ + \eta _9(\chi _2^\prime ,\chi _3^\prime )\otimes (\chi _6,\chi _7,\chi _8)\otimes (\chi _9^\prime ,\chi _{10}^\prime ,\chi _{11}^\prime ) + \eta _{10}(\chi _2^\prime ,\chi _3^\prime )\otimes (\chi _9,\chi _{10},\chi _{11})\otimes (\chi _{12},\chi _{13},\chi _{14}) \nonumber \\
&\ + \eta _{11}(\chi _4,\chi _5)_2^3 + \eta _{12}(\chi _4,\chi _5)_2\otimes (\chi _9,\chi _{10},\chi _{11})_{3_1}^2 + \eta _{13}(\chi _4,\chi _5)_2\otimes (\chi _9^\prime ,\chi _{10}^\prime ,\chi _{11}^\prime )_{3_1}^2 \nonumber \\
&\ + \eta _{14}(\chi _4,\chi _5)_2\otimes (\chi _6,\chi _7,\chi _8)_{3_1}\otimes (\chi _{12},\chi _{13},\chi _{14})_{3_1} + \eta _{15}(\chi _6,\chi _7,\chi _8)_{3_1}^2\otimes (\chi _9^\prime ,\chi _{10}^\prime ,\chi _{11}^\prime )_{3_1} \nonumber \\
&\ + \eta _{16}(\chi _9,\chi _{10},\chi _{11})_{3_1}^3 + \eta _{17}(\chi _6,\chi _7,\chi _8)_{3_1}\otimes (\chi _9,\chi _{10},\chi _{11})_{3_1} \otimes (\chi _{12},\chi _{13},\chi _{14})_{3_1} \nonumber \\
&\ + \eta _{18}(\chi _9,\chi _{10},\chi _{11})_{3_1}\otimes (\chi _9^\prime ,\chi _{10}^\prime ,\chi _{11}^\prime )_{3_1}^2 + \eta _{19}(\chi _9^\prime ,\chi _{10}^\prime ,\chi _{11}^\prime )_{3_1}\otimes (\chi _{12},\chi _{13},\chi _{14})_{3_1}^2,
\end{align}
which is written  as 
\begin{align}
w &= \mu _1\chi _1^2 + \mu _2(\chi _2^2+\chi _3^2) + \mu _3(\chi _4^2+\chi _5^2) + \mu _4(\chi _9^2+\chi _{10}^2+\chi _{11}^2) \nonumber \\
&\ + \mu _5(\chi _9^{\prime \ 2}+\chi _{10}^{\prime \ 2}+\chi _{11}^{\prime \ 2}) + \mu _6(\chi _6\chi _{12}+\chi _7\chi _{13}+\chi _8\chi _{14}) \nonumber \\
&\ + \eta _1\chi _1(\chi _2\chi _4+\chi _3\chi _5) + \eta _2\chi _1(\chi _6^2+\chi _7^2+\chi _8^2) + \eta _3\chi _1(\chi _{12}^2+\chi _{13}^2+\chi _{14}^2) \nonumber \\
&\ + \eta _4\chi _1(\chi _9\chi _9^\prime +\chi _{10}\chi _{10}^\prime +\chi _{11}\chi _{11}^\prime ) + \eta _5\{ 2\chi _2\chi _3\chi _4+(\chi _2^2-\chi _3^2)\chi _5\} \nonumber \\
&\ + \eta _6\{ \frac{\chi _2}{\sqrt 2}(\chi _7^2-\chi _8^2)+\frac{\chi _3}{\sqrt 6}(-2\chi _6^2+\chi _7^2+\chi _8^2)\} \nonumber \\
&\ + \eta _7\{ \frac{\chi _2}{\sqrt 2}(\chi _{10}\chi _{10}^\prime -\chi _{11}\chi _{11}^\prime )+\frac{\chi _3}{\sqrt 6}(-2\chi _9\chi _9^\prime +\chi _{10}\chi _{10}^\prime +\chi _{11}\chi _{11}^\prime )\} \nonumber \\
&\ + \eta _8\{ \frac{\chi _2}{\sqrt 2}(\chi _{13}^2-\chi _{14}^2)+\frac{\chi _3}{\sqrt 6}(-2\chi _{12}^2+\chi _{13}^2+\chi _{14}^2)\} \nonumber \\
&\ + \eta _9\{ \frac{\chi _2^\prime }{\sqrt 2}(\chi _7\chi _{10}^\prime -\chi _8\chi _{11}^\prime )+\frac{\chi _3^\prime }{\sqrt 6}(-2\chi _6\chi _{9}^\prime +\chi _7\chi _{10}^\prime +\chi _8\chi _{11}^\prime )\} \nonumber \\
&\ + \eta _{10}\{ \frac{\chi _2^\prime }{\sqrt 2}(\chi _{10}\chi _{13}-\chi _{11}\chi _{14})+\frac{\chi _3^\prime }{\sqrt 6}(-2\chi _9\chi _{12}+\chi _{10}\chi _{13}+\chi _{11}\chi _{14})\} \nonumber \\
&\ + \eta _{11}(3\chi _4^2\chi _5-\chi _5^3) + \eta _{12}\{ \frac{\chi _4}{\sqrt 2}(\chi _{10}^2-\chi _{11}^2)+\frac{\chi _5}{\sqrt 6}(-2\chi _9^2+\chi _{10}^2+\chi _{11}^2)\} \nonumber \\
&\ + \eta _{13}\{ \frac{\chi _4}{\sqrt 2}(\chi _{10}^{\prime \ 2}-\chi _{11}^{\prime \ 2})+\frac{\chi _5}{\sqrt 6}(-2\chi _9^{\prime \ 2}+\chi _{10}^{\prime \ 2}+\chi _{11}^{\prime \ 2})\} \nonumber \\
&\ + \eta _{14}\{ \frac{\chi _4}{\sqrt 2}(\chi _7\chi _{13}-\chi _8\chi _{14})+\frac{\chi _5}{\sqrt 6}(-2\chi _6\chi _{12}+\chi _7\chi _{13}+\chi _8\chi _{14})\} \nonumber \\
&\ + \eta _{15}(\chi _7\chi _8\chi _9^\prime +\chi _6\chi _8\chi _{10}^\prime +\chi _6\chi _7\chi _{11}^\prime ) + \eta _{16}\chi _9\chi _{10}\chi _{11} \nonumber \\
&\ + \eta _{17}\{ (\chi _7\chi _{11}+\chi _8\chi _{10})\chi _{12}+(\chi _6\chi _{11}+\chi _8\chi _9)\chi _{13}+(\chi _6\chi _{10}+\chi _7\chi _9)\chi _{14}\} \nonumber \\
&\ + \eta _{18}(\chi _9\chi _{10}^\prime \chi _{11}^\prime +\chi _{10}\chi _9^\prime \chi _{11}^\prime +\chi _{11}\chi _9^\prime \chi _{10}^\prime ) + \eta _{19}(\chi _9^\prime \chi _{13}\chi _{14}+\chi _{10}^\prime \chi _{12}\chi _{14}+\chi _{11}^\prime \chi _{12}\chi _{13})\ .
\end{align}

VEVs of $\chi_i$ must be much larger than the weak scale.
We assume that their VEVs are determined with neglecting 
supersymmetry breaking terms, i.e. $V_{\text{min}}=0$.
The $\alpha _i$ and $\alpha _j^\prime $ are very small except for
  $\alpha_6$, $\alpha_7$, $\alpha_8$,  $\alpha _{14}$,
$\alpha_2^\prime$ and  $\alpha_3^\prime$.
Then,  conditions of the potential minimum, 
$V_{\text{min}}=0$ are written  as 
\begin{align}
\eta _2(\chi _6^2+\chi _7^2+\chi _8^2)+\eta _3\chi _{14}^2=0, \nonumber \\
\eta _6(\chi _7^2-\chi _8^2)-\eta _8\chi _{14}^2=0, \nonumber \\
\frac{1}{\sqrt 6}\eta _6(-2\chi _6^2+\chi _7^2+\chi _8^2)+\frac{1}{\sqrt 6}\eta _8\chi _{14}^2=0, \nonumber \\
\frac{1}{\sqrt 6}\eta _{14}\chi _8\chi _{14}=0, \nonumber \\
\mu _6\chi_{14}=0, \nonumber \\
\eta _{14}\chi _8\chi _{14}=0, \nonumber \\
\eta _{17}\chi _7\chi _{14}=0, \nonumber \\
\eta _{17}\chi _6\chi _{14}=0, \nonumber \\
\eta _9\left (-\frac{1}{\sqrt 2}\chi _2'+\frac{1}{\sqrt 6}\chi _3'\right )+\eta _{15}\chi _6\chi _7=0, \nonumber \\
\eta _{15}\chi _6\chi _{8}=0, \nonumber \\
-\frac{2}{\sqrt 6}\eta _9\chi _3'\chi _6+\eta _{15}\chi _7\chi _8=0, \nonumber \\
\eta _{10}\left (-\frac{1}{\sqrt 2}\chi _2'+\frac{1}{\sqrt 6}\chi _3'\right )\chi _{14}=0, \nonumber \\
\mu _6\chi _6=0, \nonumber \\
\mu _{6}\chi _7=0, 
\label{appendix3}
\end{align}
where  $\chi_i$s are denoted as VEVs.
Let us consider vacuum alignments of    $(\chi_{6}, \chi_{7}, \chi_{8})$  
and $(\chi_{2}^\prime, \chi_{3}^\prime)$. Since we have a solution
\begin{align}
&\eta _{2}(\chi _6^2+\chi _7^2+\chi _8^2)+\eta _3\chi _{14}^2=0, \quad
\eta _6(\chi _7^2-\chi _8^2)=0, \quad
\frac{1}{\sqrt 6}\eta _6(-2\chi _6^2+\chi _7^2+\chi _8^2)=0,  \nonumber \\
&\eta _8=\eta _9=\eta _{10}=\eta _{14}=\eta _{15}=\eta _{17}=\mu _6=0,
\end{align}
 the vacuum alignment $\chi _6=\chi_7=\chi _8$ is a possible solution. 
On the other hand,
 $\chi_{2}^\prime=\chi_{3}^\prime$ is not guaranteed in
 Eq.(\ref{appendix3}). We may need another mechanism
 to realize the vacuum alignment  of   $(\chi_{2}^\prime, \chi_{3}^\prime)$
\cite{extra}.

\end{document}